\documentstyle[epsf]{aa}
\begin{document}
\onecolumn
\title{On correlations between diffuse interstellar bands}
\author{C. Moutou$^1$, J. Kre{\l}owski$^2$, L. d'Hendecourt$^3$, J. Jamroszczak$^2$}
\offprints{C.Moutou}
\institute{$^1$ European Southern Observatory,\\
Alonso de Cordova 3107, Santiago de Chile,\\
cmoutou@eso.org\\[0.3cm]
$^2$Toru{\'n} Center for Astronomy, \\ 
Nicholas Copernicus University, \\
Gagarina 11, Pl-87-100 Toru{\'n}, Poland \\[0.3cm]
$^3$ Institut d'Astrophysique Spatiale, CNRS,\\
Universit\'e Paris Sud,\\ 
F-91400 Orsay, France \\}
\date{Received data  ; Accepted date:}
\authorrunning{C. Moutou et al.}
\titlerunning{On correlations between DIBs}
\maketitle
\begin{abstract}
One way to better apprehend the problem of diffuse interstellar bands (DIBs) is
to search for correlations between the bands in a large sample of spectra
towards various lines of sight: a strict correlation may imply that a common
carrier is at the origin of the bands, whereas a non-correlation means that different species
are involved. We propose this observational
test for 10 DIBs collected in up to 62 Galactic lines of sight. 
Strong DIBs do not strictly correlate, and sometimes the correlation is very poor.
Only one example of a strict correlation has been found in our sample between the DIBs at 6614 
and 6196 \AA, that could significate a single carrier for those two bands.
The general absence of strict correlations is discussed in the context of
molecular carriers for the DIBs.
\end{abstract}
\newpage
\section{Introduction}
Since the first characterization of a visible diffuse band as an 
interstellar feature by Merill (1936), the observational study of DIBs
has made much progress. However, no certain DIB identification has
occured from that time (Herbig 1995 and Snow
1995). The present tendency is to gather the DIBs into subclasses, or families,
of common properties.
Chlewicki et al. (1986, 1987) and  Kre{\l}owski \& Walker (1987) have 
made the first attempts to build DIB families. Inside a family, the intensity ratios of DIBs 
should be constant (Miles \& Sarre, 1993).
At that time, only a few bands were classified, and these correlations rely on poor statistics.
A more extensive correlation study has been recently published (Cami et al.
1997).

As the idea that the carriers were molecular was progressively adopted 
(L\'eger and d'Hendecourt 1985, Crawford et al. 1985, Fulara et al. 1993), it
became clear that the observed DIB spectrum could be a mixture of vibronic states of
a large, but not infinite, number of molecular species. Therefore,
the strength ratio of any two DIBs originating in the same molecule should
be identical along any sightline. In such a case we must observe a very
tight correlation, all the scatter resulting only from the observational uncertainties.
However, an observed strict correlation is not the proof
of a single carrier, as very close species can also be invoked (i.e. species in which abundances
have exactly the same behaviour). Finally, a non-correlation between two
DIBs undoubtly implies that they are not carried by the same species.

The goal of the present paper is to use a large sample of echelle spectra acquired
at the Mc Donald and Pic du Midi Observatories, to survey the correlations between a few DIBs.
The strong DIBs at 4430, 5780, 5797, 5850, 6196, 6234, 6270, 6284, 6379 and 6614 \AA\
are surveyed in 11 to 62 lines of sight.
 
\section{The observational material}
Our observational material originates from two distinct samples, taken at Mc Donald and
Pic du Midi Observatories.

Mc Donald observations made use of the Cassegrain echelle spectrograph,
installed at the 2.1m telescope (Kre{\l}owski \& Sneden, 1993). The spectra 
are at a resolution of 60,000 over the range 5600 -- 7000\AA\
and of especially high S/N ratio -- usually 500 or more. The DIBs
at 5780, 5797, 5850, 6379 and 6614 \AA\ are measured in 50 stars, and the DIBs 6234, 6270 
and 6284 \AA\ are measured in 36 stars. 

Pic du Midi observations have been performed with the aid of the 2.03m 
Bernard Lyot Telescope in July 1995 and February 1996. 
The instrument used is the echelle spectrograph MUSICOS (Baudrand \& B\"ohm, 1992). 
Within two exposures, it
allows coverage of the whole visible range (3850-8750~\AA) at a resolving
power of 40,000.
46 orders in the ''blue'' range (3850-5400~\AA) and into 44 orders in the ''red'' range 
(5100-8750~\AA) compose the whole spectrum. We performed some redundancy with the previous
measurement sample, and added 12 lines of sight. Also the 4430 \AA\ DIB was measured
in these spectra.

To separate the orders and 
correct them for the Blaze distortion we used the data reduction software 
developped for MUSICOS by T. B\"ohm 
and J-F. Donati (private communication). The wavelength calibration is provided by a 
Thorium-Argon lamp. A tungsten lamp is used to flat-field the 
spectra. The order extraction is of good quality and allows
the easy merging of individual orders to build broader spectral ranges.
As an example, the merging of the orders near the broad 4430~\AA\
feature is shown in Fig. 1, which was obtained by multiplying the flux
in separate orders by a constant. 

Table 1 characterizes our sample of target stars. The
stars are bright (m$_v <$ 7) and nearby, increasing the probability that
they are obscured by a single cloud. A few highly reddened stars are also present,
to see how the correlation plots extrapolate to high color excesses.
However, these targets may have overlapping features which create confusion and we will rely
more on low-reddened stars, unlike most previous DIB surveys.

The reddened stars have been observed together with some standards to 
separate the interstellar features from stellar and telluric contamination.

\section{Measurements of the DIB strengths}

The central depths of the interstellar diffuse features 
considered in this paper have been measured and tabulated in Table 2.
The standard error  was estimated by multiple measurements. It originates mostly from the
continuum setting. We chose to list only the central depths (CD) of the
features because they are less sensitive to contamination than equivalent widths (EW).
For the broad features such as 4430 \AA\ the profile is free from stellar 
contamination only in the spectra of extremely hot stars. For this reason, we 
used the spectrum of the O6 star HD210839 as a typical profile.
This model spectrum was scaled to fit the observed bands, when
contaminated by stellar lines. 
The 5780 and 5797~\AA\ DIBs are measured with a low uncertainty due to their strength
(Kre{\l}owski \& Sneden 1993). 
The 5850, 6234 and 6270 \AA\ DIBs are shallow and their measurements thus suffer a 
larger uncertainty. The 6284 \AA\ band is blended by
the atmospheric $\alpha$ band of molecular oxygen which limits the precision of measurement.
The narrow 6196, 6379 and 6614 \AA\ DIBs are free of any contamination and their depth 
can be determined quite precisely. 

We checked that in most cases the bandwidth variation was negligible in the correlation   
plots. This means that the correlation in central depths reflects well the
correlation in EW. It may differ in the cases where the bandwidth is affected by special effects
such as a strong rotational broadening, named the ``rocket effect'' (Rouan et al. 
1997, Lecoupanec et al. 1999). This effect should only be observed in strongly irradiated environments
and is expected for large molecular carriers.

\section{Discussion} 
\subsection{Correlation analysis} 
We first checked the correlation degree between
our DIB measurements and $E_{B-V}$. A weak correlation is observed, which is 
usually interpreted by the differing behaviours
of DIB carriers and grains (responsible for the reddening). 
The moderate correlation between DIB intensities
and color excess is the reason why we chose to plot correlation diagrams directly
in DIB central depths CD rather than in CD/E(B-V). This also offers the advantage that 
the error bars are not enhanced by the uncertainty in E(B-V). 

For correlation plots of each DIB pair we
calculated the best fit linear curve. As an objective estimate
of the correlation level, we used the Kolmogorov-Smirnov test
(Peacock 1983).  For each point we measured the distance to the
best fitted line and we plotted the cumulative probability of
this distance to be smaller than a given value.  In the case of a
strict correlation, this curve follows the law $\int exp(-2x^2)
dx$ (Kolmogorov curve). The Pearson correlation coefficient is also 
calculated, which takes into account the error bars:\\
 $$ R =\frac{Cov(B1,B2)}{\sqrt{Var(B1) Var(B2))}}* \frac{1}{
\sqrt{(1+\frac{Var(E1)^2}{Var(B1)^2})(1+\frac{Var(E2)^2}{Var(B2)^2})}} $$ 
where B1, B2, E1 and E2 are respectively the values of the measurements
and their error bars. Error bars on the correlation coefficients are deduced 
by adding the contribution of each measurement and its variance.
 We normalised the coefficients to the
largest sample of stars (62 elements), so that it takes into
account the variability of the correlation due to the
different samples used. The correlation coefficients are listed in Table 3.
For the DIBs at 6270, 6234, and 6284 \AA, 36 lines of sight are
measured, while for the 4430 \AA\ DIB, only 11 stars are considered.
For these case, more measurements would be necessary to know the 
strength of the correlation at the same confidence level than for the other pairs of DIBs.

We observe roughly three degrees of correlation:  i) 29 couples show
a non correlation with
low coefficient (roughly R $<$ 0.7) (Fig. 2), ii) 15 couples show a rather good
correlation (0.7 $<$ R $<$ 0.95) (Fig. 3), and iii) one couple shows a strong
correlation: the pair 6196/6614 (R $>$ 0.95) (Fig. 4).\\

\subsection{Weak and medium correlations}  
The worst correlations are observed for the pairs of DIBs:  5850/5780,
6284/6379, 6614/5850, 6196/5850, 5797/6284 (some examples are shown
in Fig.  2).  For these cases, as for all correlations
where R $<$ 0.7 (Tab. 3), we can state
that the DIBs do not originate from the same
carrier and also that the carriers have distinct behaviour with 
respect to the physycal conditions. The observation that 
{\bf usually one strong DIB corresponds
to a single species} is thus for the first time confirmed on a 
strong statistical basis. Also the correlations with a weaker statistical basis 
have obviously a lower coefficient, as for the 4430 \AA\ DIB.

The second level of correlation (Fig. 3) exists for
instance between the DIBs at 5780 and 6284 \AA. The correlation coefficient
is higher, but it is obvious that a few points are more than 3$\sigma$ far away 
from the best fitted line. This degree of correlation corresponds to the 
already proposed "families" ({\it e.g.} in Kre{\l}owski and Walker 1987). 
Also, Jenniskens et al. (1993) state that 5780 and 6284 DIBs behave similarly
in Orion, and that they could both be linked to the neutral 
gas. The presence of similar substructures inside both features
was also interpreted as a possibility of a single carrier (Jenniskens and D\'esert 1993).
But later it has been remarked that some lines of sight show very discrepant band ratios 
(Kre{\l}owski and Sneden 1993).

As for this example, we observe many rather good correlations, corresponding to close species
but obviously different molecules between the DIBs at 5797/5850, 5780/6614, 5780/6196, 5797/6614,
5797/6196, 6234/6196, 6379/6196, 6614/6270, 6284/6270. 
Note that usually these pairs of DIBs are of similar bandwidths, but this is
not systematic. It is in agreement with the previous statement that broad and narrow DIBs  
define different subclasses (Herbig 1975, Jenniskens and D\'esert 1995). 
Although based on a much smaller statistical sample (11 stars), the 
correlation seems better in 4430/5780, 4430/6270, 4430/6284 (all ``broad'' DIBS) than in 4430/5797 
or 4430/5850 (5797 and 5850 \AA\ DIBs are narrow). We stress that the strong DIBs that correlate
the best with the broadest DIB at 4430 \AA\ are also quite broad.
This observation could be an indication that two distinct populations of
carriers exist (see section 4.3). \\

\subsection{One strict correlation?}
\subsubsection{Intensity ratio}
The strongest correlation found in our sample is between the two DIBs at 6614 and 6196 \AA. They are 
also of similar FWHM: in average 2.6 and 1.7 cm$^{-1}$ (1.14 and 0.65 \AA). 
Fig. 4 shows the correlation between these two DIBs. 
For this case we also show the correlation using EW measurements to prevent any width or profile effect. 
Both CD and EW plots show a very strong correlation.
Of the 62 measurements, only 4 deviate from the least-squared linear fit 
in CD, by more than 2$\sigma$.  
The correlation coefficient equals 0.97$\pm$0.14 and the cumulative probability curve is 
similar to the theoretical Kolmogorov curve (Fig. 4, bottom).
Finally, the correlation coefficient of the EW plot is 0.98$\pm$0.18.

 The lines of sight where the 6196/6614
ratio differs the most from the mean value are towards the stars HD164402, HD20041, HD43384. 
HD164402 presents strong and narrow stellar lines in its spectrum which blend interstellar features,
so that this target may be a poor candidate to measure DIBs.
HD20041 is the illuminating star of the reflection nebula VdB10 and, towards this line
of sight, the bands are suprisingly broad. This could be an evidence of the
rotational effect due to interactions between atomic and molecular gas close to the star
(Rouan et al. 1997), thus implying that the DIB carrier are molecules. 
The EW correlation plot shows that this measurement is closer to the
best fitted line, which supports this interpretation. The opposite situation is observed
for the star HD43384, towards which the DIBs are especially narrow. Also in
that case, the EW correlation plot shows a better result. Towards the star HD183143, where we observe
an effect averaged over several 
clouds, the ratio of the bands\ CD is close to the average whereas it is more
discrepant from the average relation between EW. The reason 
is the multiple convolution of bands with various velocity fields, introducing an error in the EW estimation.

We carefully considered the case of Orion stars HD37020, HD37022,
and HD37023. It has been observed previously
that the DIB strengths in this cloud were anomalously weak with respect 
to the reddening (Jenninskens et al. 1993, Kre{\l}owski and Sneden 1993).
We observe here that both 6196 and
6614 \AA\ DIBs are weak towards these stars, and the 6614/6196 
ratio is of the same order as the ratio in other targets. That means that the
radiative conditions in the Orion cloud are such that the carrier(s) of
6196 and 6614 \AA\ DIBs is(are) destroyed, either by fragmentation of its molecular skeleton or
by change of its ionisation state (Sonnentrucker et al. 1997). 

The conclusion of our observations is therefore that either the two DIBs originate from a single
molecule, or that their carriers are very strongly linked, so that their abundances
behave essentially the same regarding the interstellar conditions.

\subsubsection{DIB profiles}
In recent works based on high resolution spectroscopy of DIBs, a three peak
substructure has been found in 6614 \AA\ DIB (Sarre et al. 1995, Ehrenfreund and Foing 1996, Kerr et al. 1996). 
The 6196 \AA\ DIB has been less extensively studied, but it  
seems to have a more symmetrical profile with a single component (Kre{\l}owski 
and Schmidt 1997). Is this apparent mismatch
between the two profiles a strong point against a common carrier of the two bands?
Not necessarily. 

First we can consider the possibility that the three peak
structure of 6614 \AA\ is not an intrinsic profile, but an overlap of coincident features.
Even if the interpretation of the profile in terms of rotational contours is very appealing
in the context of molecular DIB carriers, it is nevertheless based on a sparse sample of 
heavily reddened stars.
So the possibility exists that only one component over the three peaks is 
really correlated with 6196 \AA. Our current data set is insufficient to address this issue.

Secondly, the DIB profiles depend on the internal
couplings with other vibrational levels, and thus may not be the same for
two distinct levels (Leach 1995). But this should be quite a marginal effect, because
in the diffuse medium the ground level should be the most populated, and
in any case, it would be an important test to search for profile
variations of 6196 and 6614 \AA\ DIBs in various lines of sight. 
Different profiles should indeed evolve in a similar way, if the bands have a common carrier. 
Such a critical test is  
necessary to confirm or deny the proposal of a unique carrier for the 6196 and 6614 \AA\ DIBs.

The pair of bands 6196/6614 \AA\ is thus one of the first examples of DIBs
which could be related to a single species. Another pair of DIBs proposed to be
carried by a single species are the 9577/9632 \AA\ bands, measured in only a dozen 
lines of sight and tentatively attributed to C$_{60}^+$ (Foing and Ehrenfreund
1994, 1997). However, the correlation between these two DIBs is far from
being confirmed over a strong statistical basis. Moreover,
Jenniskens et al. (1997) used a different method to
remove telluric features, and did not conclude to the identification of C$_{60}^+$.

\subsection{Comparisons with previous families}
The first attempts to search for DIB correlations were based on small samples of
stars, and aimed at observing common properties of DIBs versus color
excess (Chlewicki et al. 1986, 1987). The three ''families'' introduced by
Kre{\l}owski and Walker (1987) gather bands where the intensity ratios are
similar. Their work relies on poor statistics and thus could not evidence common
DIB carriers, as expressed by Jenniskens and D\'esert (1995). We note that 
the 6196 and 6614 \AA\ bands were not classified
in the same family in this former work, while the correlation is very tight in
our much larger target sample. Our analysis supports the loose correlation
observed by Kre{\l}owski and Walker (1987) between the pairs: 5797/5850 \AA\
(also pointed out by Josafatsson and Snow (1987) in reflection nebulae),
5797/6614 \AA, and 5850/6614 \AA. For none of these three pairs is the correlation strict.
On the other hand, we find that the 6379 \AA\ DIB is not a member of the
third family, as the correlation is weak with the other members.

A larger sample (26 stars) has been studied by Benvenuti and Porceddu (1989) who
found some weak correlations between DIBs.
Our results are in agreement with their conclusions for the bands that we
both measured. 

The classification of DIBs into narrow, broad and very broad features has already been
proposed (Herbig 1975, Jenniskens and D\'esert 1995). It is also consistent
with the "families" of Kre{\l}owski and Walker (1987).
Besides the search for strict correlations and
common carriers, we can confirm that DIBs of similar bandwidths correlate
better with each other than do bands of different width. 
Although this statement does not offer a strong constraint for laboratory investigations
on DIB carriers, it may suggest some interesting physical properties of the related species.  

A recent study of DIB correlations has been published by Cami et al. (1997)
which can be compared to our results. They
adopted a statistical way of calculating the error bars; their
correlation coefficients are therefore of different meaning. Also, they 
searched for correlations between a large number of DIBs assuming ionised carriers, and
thus explored 13 lines of sight with particular properties.
Our approach differs from theirs in the sense that we are probing the intrinsic
spectroscopic properties of the carriers, without any {\it a priori} concept of their
nature. Also our star sample is more extended and contains various types of
clouds that allows better statistics. Cami et al. do not discuss the case for
a strict correlation, between the DIBs at 6196 and 6614 \AA.
We agree with the ``close
species family'' they propose, formed by the DIBs at 5797, 6379 and 6614 \AA.
We are unfortunately unable to discuss the other families proposed by Cami et 
al. as we did not measure band intensities when the signal-to-noise
ratio was too low for reliable statistics.   

In summary, we believe that our results offer an useful complement to previous
searches for DIB families, because of the much larger sample of
stars; the comparison is limited by the reduced number of common measured DIBs.

\subsection{Implications}

Let us consider the implications of our observations, regarding
the most commonly invoked carbonaceous DIB carrier candidates: 
PAHs (polycylic aromatic hydrocarbons) and linear chains.
\subsubsection{PAHs}
PAH species have been proposed for 12 years as possible DIB carriers (L\'eger and
d'Hendecourt 1985, Crawford et al. 1985, L\'eger 1995). Laboratory studies have shown that
each open shell PAH (ions or radicals) possesses a dominant absorption band in the DIB range, and few weaker bands
as vibrational sequences (Salama et al. 1996, d'Hendecourt et al., not published). 
The vibrational progression
bands which could be correlated to the electronic origin are more likely
characterized by a separation of 50 to $\sim$ 3000 cm$^{-1}$ 
and thus we may expect such sequences
in the DIB spectrum. Unfortunately, they may be {\bf too weak} to be easily detected in a
large number of interstellar spectra, especially in moderately reddened 
stars. Some examples of very weak interstellar features related to stronger ones
have been observed in a reduced number of lines of sight (Kre{\l}owski et al. 1997).
A laboratory investigation of a derived perylene species has also led to meaningful 
observations, in terms of DIB profiles and sequences (Moutou et al. 1997).

In view of our results, we  can firmly confirm that strong DIBs are not 
correlated in most cases; we did not measure weak DIBs as our spectra were
not of sufficient signal-to-noise ratio and priority was given to a large set of targets. 

The energy 
gap between the two correlated bands at 6196 and 6614 \AA\ is 1019 cm$^{-1}$. 
If we would consider the DIB 6614 \AA\ as an electronic
ground state, then 6196 \AA\ could be an excited vibrational mode situated at 9.8 $\mu$m. 
This mode is compatible with known
molecular vibrations and corresponds to a strong mode of aromatic molecules
(not optically active in the infrared). This fact is obviously not an argument for claiming
that the two DIBs are carried by aromatic species, but the unique carrier hypothesis
is not in contradiction with the PAH hypothesis.

\subsubsection{Unsaturated chains}
Unsaturated carbon chains, as well, exhibit a strong origin followed by weaker bands
(Fulara et al. 1993, Maier et al. 1995). A mixture of photon-resistant C$_n$H$_m$ radicals and
related ions could account for a fraction of the DIB spectrum, each
species being responsible for approximately one to four bands.
For the same reason as for PAHs, our observations are compatible with the DIB carriers
being unsaturated carbon chains, and thus they do not allow to support the unsaturated
chain model nor the PAH model, if one has to be prefered.
Moreover, the number of possible linear molecules with the required 
robustness is probably too small to account for the whole spectrum, and the spectroscopy of
both kinds of molecules should be more investigated in the future.


\section{Conclusions}
From this gathering of observations and DIB measurements we can extract a few conclusions which are
of importance for the DIB classification into distinct families:
\begin{itemize}
\item
The DIBs are correlated with E$_{B-V}$ with a correlation coefficient close to 0.8;
none of the DIB intensities can be precisely predicted from the E$_{B-V}$ value.
\item
The two DIBs at 6614 and 6196 \AA\ are strongly correlated. This is
the only case in our sample where we can suggest a common carrier,
this conclusion being carried by a statistical analysis over 62 
lines of sight. However the question of their different profiles has to
be clarified by further observation at high S/N and spectral resolution, where 
we could compare the variations of intrinsic profiles or bandwidths 
with respect to the physical properties. 
\item
Some DIB pairs as
5797/5850, 6284/5780 \AA\ are well-correlated, which is the evidence
of species of close properties, but the DIBs originate from different carriers.
This supports the previous tentative classification
in narrow, intermediate and broad bands. However, this result is not 
strong as it does not constrain the laboratory search for potential carriers.
\item
The fact that in most cases the strong DIBs are not perfectly correlated
indicates that their carriers produce usually a single strong feature; other
features originating in these carriers are probably weaker. This result is compatible with
the conclusions obtained by laboratory work on aromatic and linear 
carbonaceaous species (Salama et al. 1996, Fulara et al. 1993, Leach 1995).
As a continuation of this work, we would stress the necessity of obtaining more measurements of
weaker features and then try to retrieve the vibrational sequences that can be expected in the
case of molecular carriers. \\[0.5cm]
\end{itemize}

{\it Acknowledgements :} The authors wish to express their gratitude to the 
staff members of the Pic du Midi and McDonald Observatories where the spectra 
have been acquired. Also we thank C. Catala and his colleagues for 
introducing us to the MUSICOS spectrograph and its reduction software,
and Alain L\'eger, F.X. D\'esert and M. Schmidt for many helpful discussions.
The paper is a part of the joint project sponsored by the French Embassy
and the Polish State Committee for Scientific Research (grant 2.P304.010.07), and the
bilateral PICS programme. JK wants to
express his gratitude to the Kosciuszko Foundation for supporting his stay
at McDonald.

\begin{table*}
\caption{The list of target stars is given, including: the HD number, 
spectral type and luminosity class, the reddening E(B-V), the visual 
magnitude, rotational velocity (km/s). }
\begin{tabular}{|lllcc||lllcc|}
\hline 
\small
HD&Sp T & E(B-V)& m$_v$& V$sin$i&HD&Sp T & E(B-V)& m$_v$& V$sin$i\\
\hline
&&&&&144217&B0.5 V&0.17&2.62 &130\\
2905 & B1 I  & 0.33& 4.16   & 62&144470&B1 V&0.19&3.96&142\\
5394 & B0 IV  & 0.10  & 2.47   & 300   &145502   & B3 V& 0.25&4.01&199\\
8065&A0 I&0.39&6.46&--&147084&A4 III&0.96&4.55&15\\
10516 & B2 V  & 0.17 & 4.07 & 450&147165& B2 III + O9.5 V& 0.36& 2.89& 53 \\
12953&A1 I&0.62&6.28&30&147933&B2 IV&0.44&5.02&303  \\
13267 & B5 I  & 0.41& 6.36   & 53   & 148184&B2 IV&0.48&4.42&134  \\
14489 & A2 I  & 0.40 & 5.17  & 25   & 149757   &O9.5 V &0.29&2.56 & 379 \\
20041 & A0 I&0.73 &5.79  & --& 154445   & B1V   & 0.39    & 5.64 &174 \\
21291 &B9 I & 0.42 &4.21  & 29 & 163472 & B2 IV-V  & 0.30 & 5.82 &  120 \\
 21389 & A0 Iae & 0.56 & 4.54 &6 & 164284 & B2 Ve &0.18&  4.64  &  221  \\
 22951 & B0 V & 0.23  &  4.97  &  51&164353   & B5  Ib  &  0.23 &3.97 &  22\\
 23180 & B1 III&0.26&3.82 & 85 & 164402&B0 I&0.22&5.77&100 \\
 24398 &B1 Ib&0.31 & 2.85 & 59 &166182&B2 IV&0.04&4.36&-\\
 24534 & O9.5pe& 0.56 &6.10 & 150  &166937&B8 I&0.34&3.86&54 \\
 24760 &B0.5 V + A2 V &0.06 & 2.89&153 &170740&B2 V&0.45&5.72&-\\
 24912 & O7.5 III & 0.29 & 4.02 & 216 & 179406  &  B3 V &0.31& 5.34&187 \\
 25204&B3 V + A4 IV&0.30&3.47&75&183143 & B7 Iae & 1.28& 6.86& 59 \\
 34078   & O9.5 V & 0.50 & 8.00  & 5 & 184915  & B0.5 III & 0.22& 4.95  &259 \\
 37020   & B3 V & 0.37 & 6.36 & 112& 193237& B2 pe & 0.61& 4.81 &75  \\
 37022   & B8 III & 0.96& 8.31 &127 &198478&B3 Iae&0.54&4.84 &35  \\
 37023   & O9.5 Iae &0.26 & 4.29 & 109&199478&B8 I&1.00&5.67&-\\
 37042&B1 V&0.32&6.39&-&199579&O6 Ve&0.34& 5.96 &  170 \\
 37061&B1 V&0.50&6.82&-&202850&B9 Iab&0.13& 4.23  &  28 \\
 40111&B0.5 II&0.28&4.76&130&202904&B2 V&0.10&4.43&261 \\
 41117 &B2 Ia & 0.44&4.63& 36&203064 &  O8 e & 0.28 & 5.00 &  328\\
 41335 &B2 V &0.15 &5.21 &419 & 206165   &    B2 Ib &0.46&4.73 &36 \\
 42087 &  B2.5 I &0.35 &5.75 &37& 206267   & O6f   &0.50& 5.62&  154\\
 43384& B3 I&0.58 &  6.25  &51&207198&O9 II &0.56&5.95& 76  \\
 45725& B3 V  &0.08 &4.60  &346&207260& A2Ia   &0.50 &  4.29 &33 \\
 48099& O6 e  & 0.34 &6.37 & -- &208501 & B8Ib&0.75 & 5.80     &  53 \\
 47129& O8 V&0.33 &6.06  &80  &209481&O9 V&0.34&5.56 &130 \\
 47240&B1 I &0.33&6.15 & 126&209975&O9 I&0.33&5.11 &33 \\
 54662&O7 III& 0.27&6.21 &91&210839 &O6 If& 0.46&5.04 &  285 \\
 89353&B9.5 I&  0.24 &5.34 &-- & 212076&B2 IV&0.08&5.01&134  \\
141637 & B3 V  & 0.13 &4.64 &300 &213420&B2 IV&0.16&5.95 & --\\
142096&B3 V&0.20&5.02&207&216200&B3 IV& 0.25 & 5.92& 225 \\
142114 &B2.5 V&0.12&4.59&308&218376&B0.5 IV  & 0.21 &4.85 &  50   \\
142184 &B2.5 V&0.15&5.42&349&223128&B2 IV&0.16&5.95&-- \\
143275 &B0.5 IV&0.12&2.32&181&224572&B1 V &0.16&4.88&189  \\
\hline
\end{tabular}
\end{table*}
\normalsize
\begin{table*}
\caption{ List of DIB measurements. For each star of the sample,
the central depths and standard errors are given in \%. The uncertainty 
(in parenthesis) is estimated by multiple
measurements. ``-'' means that the measurement is not possible, 
either because of band weakness or lack of data.}
\begin{tabular}{|l|cccccccccc|}
\hline 
\small
HD&4430& 5780&5797&5850&6196&6234&6270&6284&6379&6614\\
\hline
2905 && 13.4  (0.3)  &  8.5  (0.4) & 2.3  (0.1)  &  7.0  (0.5) &-&-&-&8.6  (0.3)  & 11.9 (0.5) \\ 	
5394 & & 1.8  (0.2)  &  1.6 (0.3)  & 0.6  (0.2)  & 0.9  (0.4) &-&-&- &6.5  (0.4)  &  1.2 (0.8) \\ 
14489&  & 14.7 (0.4)  &  7.3  (0.3)&  3.0 (0.1)  &  5.0  (0.4)&-&-&- & 5.6  (0.4)  &  9.7  (0.3) \\  
20041 & & 20.2  (0.4)  & 13.0  (0.9) & 4.6  (0.3)  &  9.3  (0.8)&-&-&-&  12.5 (1.3)& 22.4  (1.9) \\  
21291 & & 9.9  (0.2)  &  7.4  (0.4)  &3.1  (0.2) &  4.9  (0.5) &-&-&- & 5.6  (0.2)  &  9.6 (0.5) \\  
21389 & & 19.0  (0.3)  &  8.9  (0.6) & 3.1  (0.2)  &  7.4  (0.5) &-&-&-& 7.6  (0.9)  & 15.0  (0.6) \\  
23180 & & 4.2  (0.2)  &  8.1  (0.6) & 4.6  (0.2) & 3.4  (0.2) &-&-&-&7.0  (0.8)  &  4.5  (0.3) \\	  
22951 & & 5.0  (0.2)  &  4.4  (0.2) & 1.9  (0.1) & 3.8  (0.2) &-&-&-&3.2  (0.3)  &  5.9  (0.3) \\ 	 
24398 & & 5.0  (0.3)  &  7.4  (0.6) & 3.1  (0.4) & 4.1  (0.1) &-&-&-&8.9  (0.8)  &  6.2  (0.2) \\ 	  
24912 & & 9.4  (0.3)  &  5.1  (0.3) & 2.1  (0.1) & 4.6  (0.2) &-&-&-&5.1  (0.7)  &  7.4  (0.5) \\ 	  
37020 & & 2.4  (0.4)  &  1.1  (0.3) & 1.0  (0.2) & 1.3  (0.4) &-&-&-&1.3  (0.4)  &  1.3  (0.6) \\ 
37022 & & 3.0  (0.4)  &  1.0  (0.3)  & 0.7  (0.4) & 1.3  (0.3) &-&-&-&1.1  (0.3  &  1.2  (0.3) \\ 	  
37023 & & 2.8  (0.4)  &  1.4  (0.3) & 1.0  (0.2) & 1.4  (0.2) &-&-&-&1.4  (0.2)  &  1.3  (0.1) \\ 
34078 && 8.1  (0.4)  &  5.6  (0.3) & 2.3  (0.3) & 3.7  (0.3) &-&-&-&2.2  (0.3)  &  4.8  (0.5) \\ 
41117 & & 16.0  (0.2)  & 14.0  (0.6)  &  5.6  (0.3) & 8.3  (0.3) &-&-&-&16.3 (1.7)  & 15.6  (0.7) \\  
42087 &&  11.7  (0.3)  & 10.6  (0.8)  &  5.6  (0.3) & 6.5  (0.4)  &-&-&-& 7.6  (0.3)  & 11.5  (0.9) \\  
43384 && 20.7  (0.5)  & 14.1  (0.9)  &  5.0  (0.2)  & 11.2  (1.1) &-&-&-&  12.1  (1.5)  & 19.1 (1.2) \\  
41335 & & 1.4  (0.1)  &  1.3  (0.1)  & 0.8  (0.1)  &  2.1  (0.3)  &-&-&-& 1.3  (0.2) & 3.2  (0.3) \\
89353 &&  1.7  (0.3)  &  1.3  (0.1)  &  1.1  (0.2)   &  1.0  (0.2)  &0.5 (0.2)&1.3 (0.1) &1.2 (0.2)  & 1.1  (0.2)   &  0.7  (0.3) \\
141637 & & 4.2  (0.1)   &  1.2  (0.1)   &  1.1  (0.3)   &  1.8  (0.1) &0.7 (0.1)&1.9 (0.2)&3.7 (0.4) & 1.6  (0.1)  &  2.1  (0.3) \\ 
142114 && 3.5  (0.2)   &  1.2  (0.1)   & 0.7  (0.1)   &  1.7  (0.1)  &0.4 (0.1)&1.4 (0.1) &3.1 (0.5)  & 1.5  (0.1)   &  1.4  (0.1) \\
143275 & & 9.8  (0.3)   &  2.0  (0.1)  & 1.0  (0.1)   &  1.9  (0.2) &0.5 (0.1) &1.2 (0.1) &3.0 (0.3)  & 1.6 (0.1)   &  1.5  (0.1) \\
144470 & & 0.9  (0.1)  &  3.2  (0.2) & 1.3  (0.2)  &  3.9  (0.3)  &1.3 (0.1) &2.1(0.3)&4.6 (0.2)& 4.2  (0.1)  &  6.4  (0.3) \\
145502 &5.7(0.6) & 8.5  (0.2)  &  3.8  (0.2)  &  1.5  (0.1)  &  3.2  (0.1)  &1.3 (0.2) &1.1 (0.3)&5.7 (0.2) & 4.6  (0.2)  &  6.0  (0.3) \\  	   
147165 & & 12.0  (0.5)  &  3.6  (0.1)  &  1.6  (0.3)  &  3.7  (0.1) &1.6 (0.2) &1.3 (0.3)&5.1 (0.2) & 4.0  (0.2)  &  5.7  (0.3)\\  
147933 & & 10.1  (0.2)  &  6.7  (0.1)  &  2.9  (0.5)  &  3.4  (0.3) &1.2 (0.2) &1.2 (0.2)&3.6 (0.4) & 4.0  (0.1)  &  5.2  (0.3) \\ 
148184 & & 5.5  (0.3)  &  5.3  (0.2)  &  3.4  (0.2)  &  2.4  (0.2) &1.2 (0.2) &1.1 (0.2)&2.8 (0.3) & 4.2 (0.2)  &  4.2  (0.5) \\ 
149757 &3.7 (0.4) & 3.7  (0.1)  &  4.1  (0.1)  &  1.9  (0.1)  &  2.5  (0.1) &1.0 (0.2) &1.1 (0.1)&2.0 (0.1) & 3.5 (0.2)  &  3.4  (0.1) \\  	   	
154445 &4.7(0.5) & 9.7  (0.3)  &  7.9  (0.5)  &  2.8  (0.2)  &  5.7  (0.5) &1.8 (0.2) &3.1 (0.3)&4.1 (0.3)  & 7.8  (0.2)  & 11.8  (0.9) \\ 	   	
164402 & & 7.2  (0.4)  &  4.4  (0.2)  &  2.6  (0.3)  &  4.0  (0.4) &1.2 (0.1)&1.6 (0.1)&3.0 (0.1) & 3.8  (0.1)  &  4.0  (0.5) \\ 
166937 & & 12.3  (0.5)  &  9.2  (0.6)  &  3.8  (0.5)  &  5.7  (0.5) &1.7 (0.2)&3.3 (0.2)&8.1 (0.2)& 8.3 (0.2)  &  9.4  (0.4)\\ 
179406 & & 8.2  (0.3)  &  8.4  (0.6)  &  4.1  (0.2)  &  4.8  (0.1) &1.3 (0.4)&2.8 (0.2)&2.7 (0.2) & 9.7  (0.4)  &  9.0  (0.4) \\  
184915 &5.9 (0.6) & 7.6  (0.2)  &  3.1  (0.2)  &  1.1  (0.1)  &  4.4  (0.1)&0.8 (0.2)&2.6 (0.3)&3.6 (0.3) & 3.8  (0.2)  &  6.9  (0.6) \\	   
198478 &6.7(0.7) & 14.2  (0.9)  &  9.0  (0.6)  &  3.6  (0.2)  &  7.0  (0.2)&2.1 (0.1)&3.9 (0.4)&3.9 (0.5) & 9.5  (0.1)  & 12.4  (0.6) \\	  
199579 & & 5.4  (0.3)  &  5.5  (0.5) & 2.6  (0.2  &  3.2  (0.2)&2.0 (0.3)&1.2 (0.1)&3.68 (0.4)& 2.9  (0.3)  &  4.8  (0.6) \\
203064 & & 7.7  (0.3)  &  5.5  (0.5) & 2.4  (0.4)  &  3.7  (0.2)&1.6 (0.2)&2.0 (0.2)&6.0 (0.6) & 2.9  (0.2)  &  6.7  (0.6) \\
206165 &5.9 (0.6) & 9.5  (0.9)  &  8.9  (0.2) & 3.8  (0.4)  &  5.6  (0.2)&1.4 (0.1)&2.9 (0.3)&6.8 (0.7)& 8.4  (0.1)  & 10.5  (0.5) \\
206267 & & 10.3  (0.2)  &  9.9  (0.4) & 4.5  (0.3)  &  6.2  (0.5)&2.1 (0.3)&3.1 (0.3)&7.1 (0.7) & 6.2 (0.3)  & 11.3  (0.5) \\
207260 &5.9 (0.6) & 14.8  (0.4)  & 11.0  (0.9) & 4.0  (0.2)  &  7.0  (0.7)&1.8 (0.2)&4.9 (0.5)&9.3 (0.9) & 11.5 (0.2)  & 13.6  (0.8) \\  
208501 &5.2 (0.5) & 10.9  (0.3)  & 11.0  (0.9) & 3.8  (0.3)  &  6.5  (0.4)&2.0 (0.3)&3.5 (0.4)&6.7 (0.8) & 8.5 (0.3)  & 11.8  (0.5) \\	   
210839 & & 11.4  (0.4)  &  9.6  (0.6) & 3.1  (0.2)  &  7.4  (0.4) &1.9 (0.1)&5.2 (0.5)&7.5 (0.8)& 8.4 (0.1)  & 14.3  (0.4) \\
209975 & & 11.8  (0.3)  &  8.7  (0.2) & 3.3  (0.2)  &  6.2  (0.2)&2.0 (0.1)&3.8 (0.4)&8.0 (0.8) & 7.2  (0.1)  & 11.6  (0.7)\\
54662  && 9.5  (0.3)  &  6.3  (0.3)  &  2.8  (0.3)  &  4.5  (0.3)&1.6 (0.5)&3.3 (0.4)&8.8 (0.9) & 5.2  (0.5)  &  8.8  (0.6) \\
207198 & & 11.1  (0.2)  & 16.0  (1.2)  &  8.5  (0.3)  &  6.7  (0.2)&2.7 (0.1)&3.9 (0.4)&7.9 (0.8)& 11.8  (0.3)& 12.6  (0.9) \\
209481 & & 8.5  (0.4)  &  6.8  (0.8)  &  2.9  (0.4)  &  5.1  (0.1) &1.5 (0.4)&3.5 (0.4)&6.8 (0.7)& 11.8 (0.4)& 9.6  (0.7)\\
24534  &2.5 (0.3)& 4.6  (0.2)  &  7.5  (0.4) &  4.4  (0.6)  &  3.8  (0.2) &1.8 (0.3) &2.0 (0.2)&3.5 (0.5)&  7.4  (0.3) & 5.8  (0.5) \\	 
212076 & & 1.7  (0.4) & 1.6  (0.2) & 0.8  (0.2) &  1.3  (0.1) &0.8 (0.2)&2.0 (0.2)&2.0 (0.2)& 1.5  (0.2) & 1.3  (0.2)\\
218376 & & 5.9  (0.3) & 4.8  (0.2) & 1.8  (0.2) & 3.6  (0.2) &1.1 (0.2)&2.9 (0.3)&5.1 (0.5) & 5.5  (0.3)&  6.5  (0.4)\\
223128 & & 6.4  (0.4)& 6.3  (0.3) & 2.4  (0.4) & 3.8  (0.2) &1.1 (0.2)&2.8 (0.3)&4.8 (0.5)& 5.5  (0.2)&  6.6  (0.3)\\
47129  && 8.2  (0.4)& 5.1  (0.3) & 1.8  (0.5)& 4.2  (0.2) &0.6(0.2)&5.0(0.5)&6.7(0.7) & 5.1  (0.2) &  8.1  (0.5)\\
47099  && 8.8  (0.4) & 0.1  (0.1) & 5.4  (0.3) & 3.9  (0.2)  &-&-&-&3.4  (0.5) & 8.7  (0.4) \\
37061 & & 6.8  (2.0) & 1.6  (0.1)  & 0.9  (0.2) & 1.6  (0.4)  &-&-&-& 1.8 (0.5) & 1.6  (0.1)\\
37042 & & 1.9  (0.1)  & 0.9  (0.2)  & 0.7  (0.1)  & 0.9  (0.2)  &-&-&-&1.8  (0.1) & 1.8  (0.2)\\
47240  && 11.2  (0.9) & 8.4  (0.6) & 3.4  (0.3) & 5.7  (0.6)  &-&-&- &7.4  (0.7) &10.9  (0.9)\\
224572 & & 3.6  (0.2) & 2.8  (0.2) & 1.5  (0.3) & 2.2  (0.2)  &-&-&-&1.7  (0.2) & 4.0  (0.3) \\
213420 & & 4.1  (0.1)   & 2.6  (0.1)   & 1.6  (0.1)   & 1.9  (0.2)  &-&-&-&1.7  (0.1)   & 3.5  (0.2)  \\
216200 & & 4.6  (0.5)   & 5.3  (0.5)   & 2.6  (0.2)   & 2.2  (0.1)    &-&-&-&2.7  (0.1)   & 4.3  (0.1)  \\
144217 &7.7 (0.8)  & 8.4  (0.2)   & 2.2  (0.1)    & 0.7  (0.2)   & 3.0  (0.2) &-&-&-& 2.5  (0.3)  &  5.3  (0.6)  \\	
142184 & & 3.8  (0.1)   & 2.2  (0.1)   & 1.1  (0.3)   & 1.9  (0.4)   &-&-&-&2.2  (0.1)   & 2.3  (0.1)  \\
24760  && 3.5  (0.5)  & 1.8  (0.1)    & 0.8  (0.1)  &  2.1  (0.3)    &-&-&-&1.3  (0.2)   & 3.2  (0.3) \\
13267  && 11.6  (0.2)   & 6.5  (0.3)  &  2.3  (0.2)   & 4.9  (0.2)    &-&-&-&6.5  (0.6)   &10.6  (0.8) \\
183143 &15.4(0.9)  & 34.1  (0.9)   &20.3  (1.5)  & 7.0  (0.5)  & 14.6  (0.9) &-&-&-& 15.7 (1.5)  & 29.2  (0.9)  \\
\hline
\end{tabular}
\end{table*}
\normalsize
\newpage
\begin{table}
\caption { Correlation coefficients: $^a$ : sample
of 62 stars, $^b$ : sample of 36 stars, $^c$ : sample of 11
stars.} 
\begin{tabular}{|c||c|c|c|c|c|c|c|c|c|}
\hline
\small
&4430$^c$&5780&5797&5850&6196&6234$^b$&6270$^b$&6284$^b$&6379\\
\hline
6614$^a$&0.35 (0.015)&0.93 (0.12)&0.91 (0.13)&0.77 (0.14)&0.97 (0.14)&0.62 (0.15)&0.70 (0.08)&0.67 (0.08)&0.88 (0.13)\\
6379$^a$&0.26 (0.02)&0.78 (0.09)&0.92 (0.11)&0.84 (0.14)&0.90 (0.11)&0.60 (0.12)&0.58 (0.05)&0.54 (0.05)&\\
6284$^b$&0.39 (0.02)&0.69 (0.06)&0.59 (0.05)&0.47 (0.06)&0.67 (0.06)&0.48 (0.10)&0.71 (0.07)&    &\\
6270$^b$&0.38 (0.03)&0.66 (0.06)&0.61 (0.05)&0.47 (0.06)&0.70 (0.06)&0.45 (0.10)&    &    &\\
6234$^b$&0.22 (0.02)&0.52 (0.11)&0.67 (0.13)&0.65 (0.16)&0.62 (0.13)&    &    &    &\\
6196$^a$&0.36 (0.02)&0.93 (0.10)&0.92 (0.11)&0.79 (0.12)&    &    &    &    &\\
5850$^a$&0.22 (0.02)&0.68 (0.10)&0.93 (0.14)&    &    &    &    &    &\\
5797$^a$&0.29 (0.01)&0.83 (0.09)&    &    &    &    &    &    &\\
5780$^a$&0.39 (0.02)&    &    &    &    &    &    &    &\\
\hline
\end{tabular}
\end{table}
\normalsize
\newpage
\vskip2cm
\begin{figure}
\epsfxsize=12cm
\hbox{\epsffile{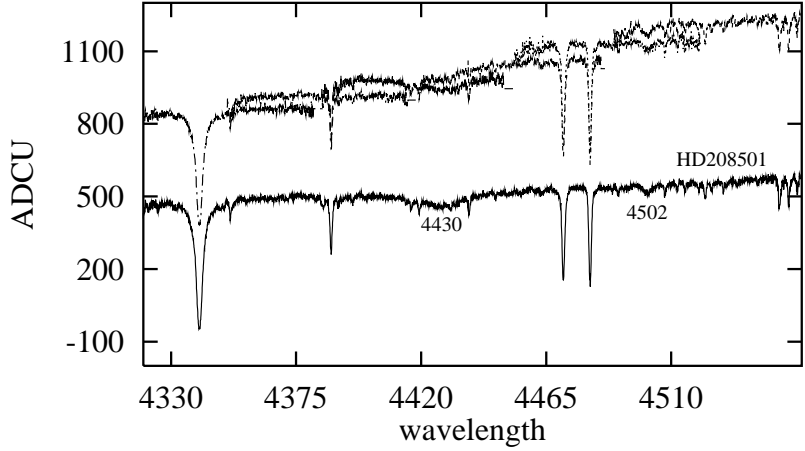}}
\caption {The spectrum of HD208501 containing the 4430 and 4502 \AA\ diffuse 
bands, obtained with MUSICOS in July 1995. Wavelengths are in \AA, while 
flux values are in arbitrary units.
The six orders (top) are merged into one spectrum (bottom).
Note the flatness of the continua in individual orders. Narrow lines are of
stellar origin.}
\end{figure}

\begin{figure}
\epsfxsize=12cm
\hbox{\epsffile{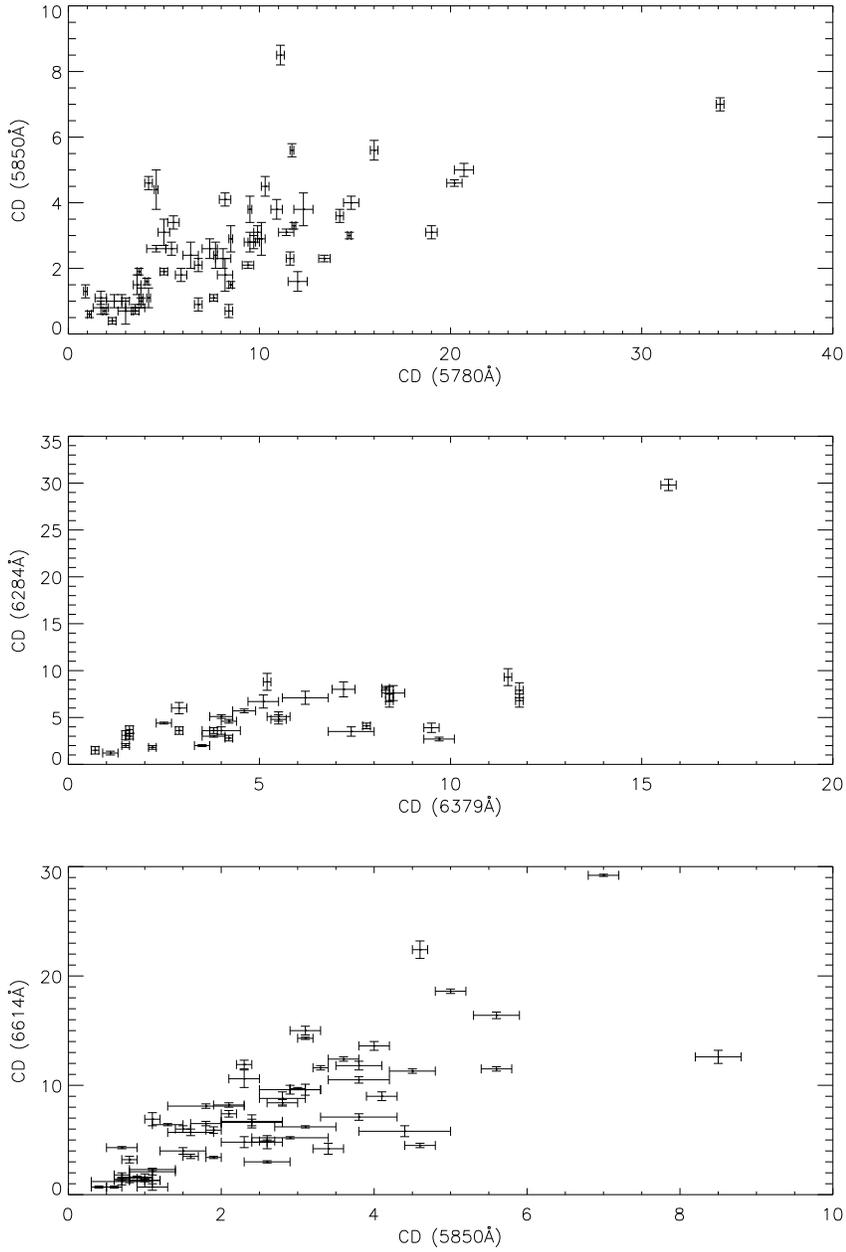}}
\caption { The correlation plots between some pairs of measured DIBs, when the 
correlation is very poor. The high dispersion indicates different
origins. CD are in \% of the continuum.}
\end{figure}

\begin{figure}
\epsfxsize=12cm
\hbox{\epsffile{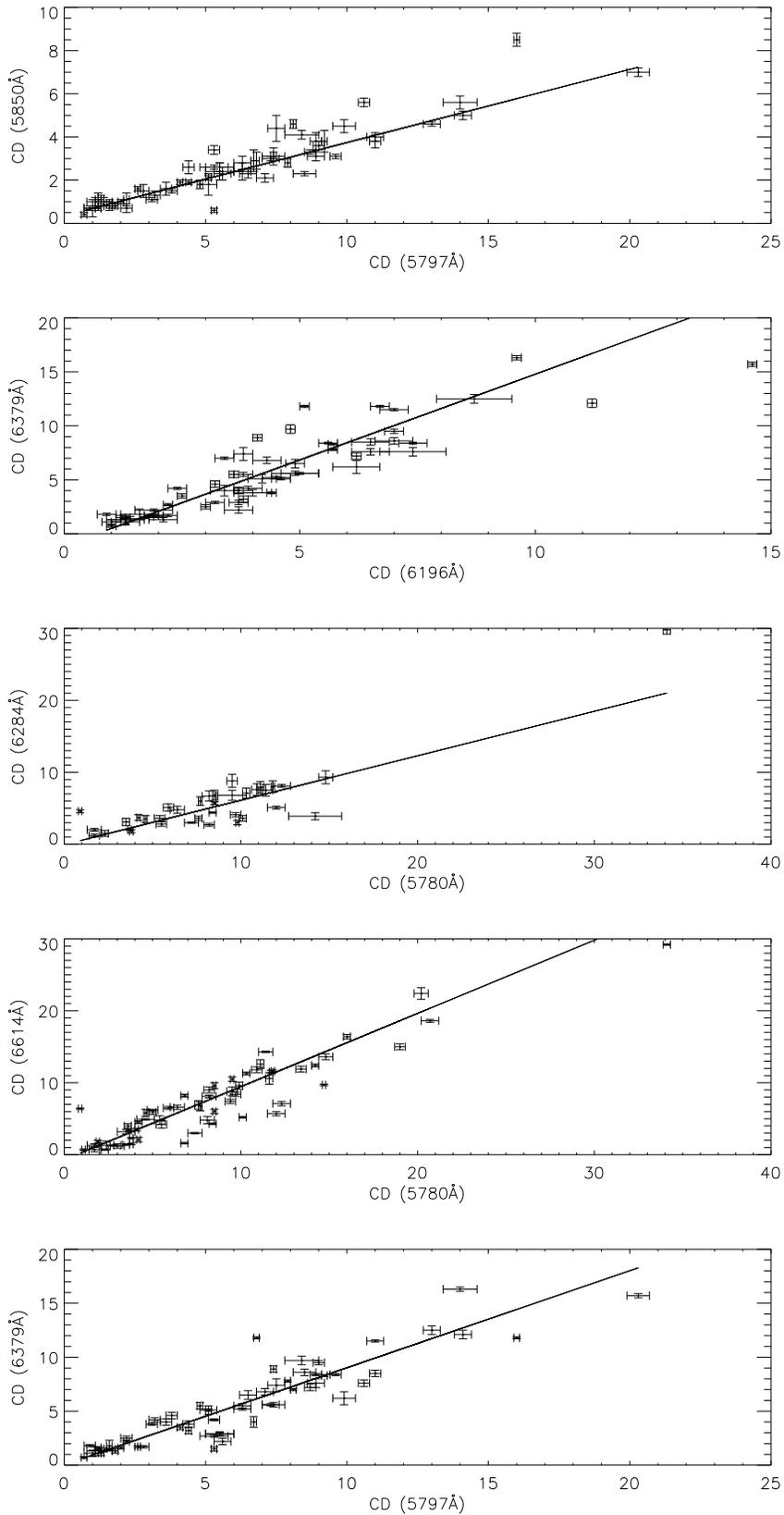}}
\caption {The correlation plots between some pairs of measured DIBs, when the correlation is
rather good.
The least-squared linear curve is also plotted.
In each plot some points obviously deviate from the fit by more than
3$\sigma$, excluding the possibility of a common carrier.}
\end{figure}

\begin{figure} 
\epsfysize=12cm
\hbox{\epsffile{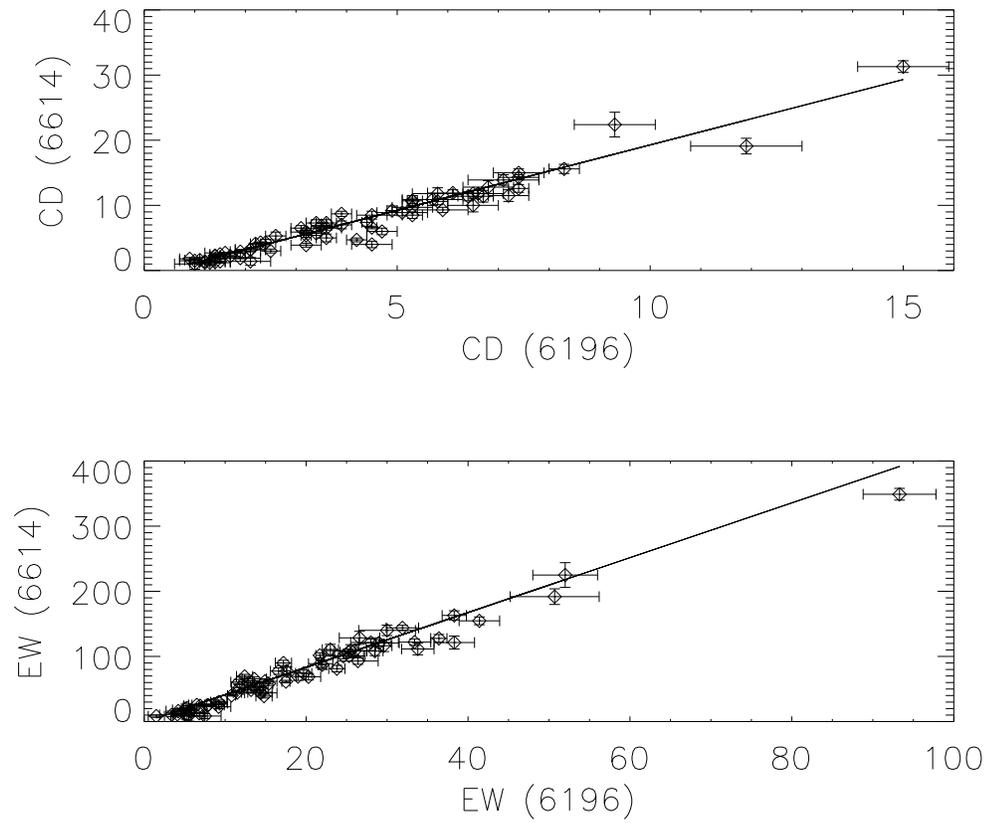}}
\caption {The correlation plot between DIBs at 6614 and 6196 \AA\ in 62
lines of sight. We show both the central depth (CD in \%) and the equivalent width (EW in \AA)
correlation plots. The difference between the two plots is weak and concerns mostly the most
reddened objects for which the bandwidth is the most sensitive to averaging between
different clouds. The EW correlation nevertheless shows a lower dispersion. }
\end{figure}   

\end{document}